\documentstyle[preprint,aps,prb,epsf,epsfig]{revtex}
\draft
\tightenlines

\begin{document}

\title
{\large Connective neck evolution and conductance steps in ``hot''
point-contacts.}

\author
{\mbox{A. Halbritter$^{(a)}$}, \mbox{Sz. Csonka$^{(a)}$},
\mbox{O.Yu. Kolesnychenko$^{(b)}$}, \mbox{G. Mih\'aly$^{(a)}$},
\mbox{O.I. Shklyarevskii$^{(b,c)},$\mbox{ and H. van
Kempen$^{(b)}$}} }

\address {$^{a}$Department of Physics, Institute of Physics,
Budapest University of Technology and Economics, 1111 Budapest,
Hungary} {\address{$^{b}$ Research Institute for Materials,
University of Nijmegen, Toernooiveld 1, NL--6525 ED Nijmegen, the
Netherlands}

\address{$^{c}$ B. Verkin Institute for Low Temperature Physics \&
Engineering,\\ National Academy of Science of Ukraine, 47 Lenin Av.,
61164, Kharkov, Ukraine} \maketitle

\vspace{-5mm}

\begin{abstract}
Dynamic evolution of the connective neck in Al and Pb mechanically
controllable break junctions was studied during continuous approach
of electrodes at bias voltages $V_b$ up to a few hundred mV. A high
level of power dissipation ($10^{-4}$ -- $10^{-3}$~W) and high
current density ($j\,\gtrsim \, 10^{10}$~A/cm$^2$) in the
constriction lead to overheating of the contact area,
electromigration and current-enhanced diffusion of atoms out of the
``hot spot''. At a low electrode approach rate ($\sim 10 - 50$~pm/s)
the transverse dimension of the neck and the conductance of the
junction depend on $V_b$ and remain nearly constant over the
approach distance of 10 -- 30 nm. For $V_b\,>\,300$~mV the
connective neck consists of a few atoms only and the quantum nature
of conductance manifests itself in abrupt steps and reversible jumps
between two or more levels. These features are related to an ever
changing number  of individual conductance channels due to the
continuous rearrangement in atomic configuration of the neck, the
recurring motion of atoms between metastable states, the formation
and breaking of isolated one-atom contacts and the switching between
energetically preferable neck geometries.
\end{abstract}

\section{Introduction}

Electron transport through nanometric-sized systems has been a
subject of intensive research over the past decade not only because
of its obvious technological importance, but primarily due to
fascinating effects of fundamental interest related to the quantum
nature of the conductance through a few atom metallic contacts.

These phenomena were sought mostly in the single $s-$valence metals
using the mechanically controllable break junction (MCBJ) technique
in point-contacts of submesoscopic size \cite{cris,krans,nat,cris2}
and nanowires (long connective necks between the tip and the sample)
fabricated using  a scanning tunneling microscope (STM).
\cite{pas,agr,ole,unt} More recently this effect was investigated in
contacts between two crossed vibrating wires \cite{costa},
gold-plated wafer and gold pin \cite{land} and even in commercial
and home-built relays. \cite{yas,han} In all these techniques  a
metallic contact is formed by pressing two electrodes together.
During the subsequent separation of the electrodes the conductance
decreases in abrupt steps with plateaus which in some instances are
close to integer multiples of the fundamental conductance unit
$G_0\,=\,2e^2/h$.

Individual traces of the conductance versus the electrode
displacement $G(z)$ are irreproducible owing to the different
dynamical evolution of the connective necks during the break.
Statistical analysis of the experimental data includes construction
of conductance histograms based on hundreds or even hundreds of
thousands of curves. \cite{CK} However, with the exception of
impressive experiments showing conductance quantization in lithium,
sodium and potassium point contacts, \cite{nat,thesis} the
histograms basically reflect the existence of stable neck geometries
emerging during the break. This is supported by simultaneous
measurements of force and conductance in Au nanowires, demonstrating
that jumps in $G(z)$ curves during the deformation of the connective
necks are always correlated to mechanical force relaxations and thus
to atomic rearrangements. \cite{rubio}

Since atomic rearrangement plays a key role one would expect a
perceptible dependence of conductance curves on temperature. Rather
surprisingly early STM measurements of Sirvent {\it et al.}
\cite{sirvent} carried out at temperatures of 4, 77 and 300 K
produced histograms with practically no difference in shape and
position of conductance peaks. On the other hand the only
measurements performed on Au and Cu MCBJ at room temperature
\cite{cris2} discovered more prominent structure in the histograms
than at 4.2 K. It was explained by the fact that at higher
temperatures the connective neck reaches an energetically favorable
geometry more easily. This discrepancy might be caused by the high
retraction speed of the tip in STM measurements ($10^2 - 10^4$ nm/s)
which is 4-6 orders of magnitude larger than the electrode approach
rate in MCBJ experiments.

 Room temperature measurements with MCBJ impose heavy
demands on temperature stability of the setup and require an ultra
high vacuum environment to avoid contamination or oxidation of
freshly broken surfaces.  In our experiments we combined the
cleanliness and stability of operating a MCBJ at 4.2 K with
sufficient mobility of atoms at the contact area to relax to a low
energy geometry. To this end we took advantage of the fact that
connective necks can be locally overheated at high bias voltages,
$V_b$. As the applied voltage drops mainly around the centre of the
contact within a distance comparable to the contact diameter, $d$
the magnitude of the heating is determined by the ratio between $d$
and the diffusive length for electrons, $L_i$. The latter can be defined
as $L_i=\frac{1}{3}\cdot\sqrt{l_e \cdot l_i}$, where $l_e$ is the
elastic mean free path, and $l_i=v_F\tau_i$ is the inelastic mean
free path of the electron-phonon scattering. \cite{yan} If $L_i$ is
much smaller than the neck diameter ($L_i \ll d$), the power
dissipation takes place in a volume of $d^3$ around the centre of
the junction. In the other limiting case ($L_i \gg d$) the power is
always dissipated in a volume of $L_i^3$ around the junction,
regardless of the contact diameter. According to these
considerations, and assuming the validity of the Wiedemann-Franz
condition, one can make estimations for the temperature of the
junction in the different regimes of point contacts. In the thermal
limit ($l_e,L_i \ll d$) the temperature in the  centre of the
contact  does not depend on $d$ and is given as follows:
$T_{pc}^2=T_{bath}^2+V_b^2/4\mathcal{L}$, where $\mathcal{L}$ is the
Lorentz number. \cite{holm} (Or $eV_b\;\approx\;3.63\;kT_{pc}$ at
$T_{bath}$ = 4.2 K and high $V_b$.\cite{yan})  Practically it means
that a voltage bias of $200 - 300$~mV could heat the contact up to
$T_{pc}=600 - 1000$~K as it was demonstrated for point-contacts
between ferromagnetic metals. \cite{verkin} On the other hand in the
spectroscopic regime ($l_e,L_i \gg d$) the heating of the contact is
negligible, and $T_{pc}\approx T_{bath}$. In the intermediate case
($l_e < d < L_i$) simple calculations show, that the crossover
between the ballistic and thermal limit is mainly determined by the
ratio of $d$ and $L_i$, namely: $T_{pc}^2=T_{bath}^2+d/L_i \cdot
V_b^2/4\mathcal{L}$. Even though in nanometric point contacts $l_e$
is assumed to be in the range of a few nm (due to scattering on
defects at the interface of the electrodes), at low temperature and
bias voltage $L_i$ remains much larger than the contact diameter due
to the large inelastic mean free path. Therefore, the heating is
considerably reduced compared to the thermal limit. However, $l_i$
drops to $\lesssim$ 10 nm at bias voltage $eV_b$ exceeding the Debye
energy of phonons due to the strong electron-phonon coupling. The
energy dependence of $l_i$ can be determined from the experimental
point-contact spectra of the electron-phonon interaction.
\cite{atlas} The inelastic mean free path continues to decrease at
higher energies due to the relatively large probability of
multiphonon scattering.\cite{yan} This is in agreement with
Ref.~\onlinecite{pers} where the inelastic mean free path of a hot
electron with excess energy of 100 meV was estimated to be 1 -- 10 nm
for most metals. With all the parameters being the same order of
magnitude ($L_i \approx l_e \approx d$) it is extremely difficult to
give an accurate estimation for the contact temperature, but
according to the above considerations contacts of submesoscopic size
might be heated well above room temperature with a voltage bias of a
few hundred mV.

In this paper we present conductance curves for Al and Pb
point-contacts measured at $V_b$ up to 0.4 V using the MCBJ
technique, and we give a simple analysis for the observations. These
$sp-$like metals are the subject of current interest since the
transport properties of one-atom Al or Pb contacts are determined by
the existence of $three$ conducting channels with nonvanishing
transmission and by a strong dependence of the transmission
coefficients on the degree of elastic deformation of the neck.
\cite{scheer,cuevas,scheernat,alpb,kob} In the light of recent
attempts to study  nonlinear effects in conductance in connective
necks of a few atoms  at $eV_b$ comparable to the Fermi energy
\cite{itakura} the details of the neck evolution at high $V_b$ are
of special importance.

\section{Experiment}

The basic concepts of MCBJ method were developed by Moreland and
Ekin, \cite{mor} refined later by Muller and co-authors \cite{cris3}
and described elsewhere. \cite{cris,krans,nat,cris2} In our
experiments we used a slightly modified version of the traditional
sample mount presented in the inset in Fig.~1. It includes two
additional pieces ($5 \times 3 \times1$~mm$^3$) of fiberglass board
plate placed underneath the anchoring points of the 200 $\mu$m thick
polycristalline sample wire. This increases the effective thickness
of the bending beam and thus enlarges the ratio of the electrode
displacement $\Delta_z$ to the vertical transfer of the piezodriver
$\Delta_y$. Such a modification practically does not affect the
stability of the break-junctions and gives an opportunity to cover
the range of $\Delta_z$ up to 20 -- 30 nm in one run of the
piezodriver.

Most of the measurements were performed at 4.2 K in He exchange gas
at 760 Torr to avoid effects related to thermal expansion of the
unglued sections of electrodes. Adsorbtion of He on the electrode
surface results in a giant increase of the local work function of
metals \cite{phi} and a strong deviation of tunnel current versus
electrode separation $I(z)$ curve from exponential behavior.
\cite{he} As a consequence the standard calibration procedures of
the relative displacement of the electrodes (using the $I(z)$
dependence in tunneling regime or field emission resonance spectra
\cite{rsi}) are rather inaccurate. We can estimate $\Delta_ z$ and
the approach rate $S\,=\,\Delta_z/t$ with a precision of 25 -- 30\%.
Fortunately this inaccuracy has little or no effect on our results
which are qualitatively the same in a wide range of $S$ from 5 to 50
pm/s.

Contact conductance was measured using a current to voltage
converter with a gain of 1~V/mA. Data were recorded with a Keithley
182 nanovoltmeter in a slow (1 -- 10 pt/s) mode and voltage range
from 1 $\mu$V to 10 V to cover the transition from tunneling to
direct conductance. Simultaneously a AT-MIO-16XE-50 National
Instruments data acquisition board was used in a fast (250-1000
pt/s) mode in 10 mV - 10 V range. Some fragments of conductance
curves were measured at the rate of 10000 pt/sec for 1-2 minutes.

All results presented below are based on a careful analysis of a few
hundreds $G(z)$ traces obtained on 5-6 different samples for each
metal.

\section{Results and discussion}

In this section we present experiments on locally overheated Pb and
Al MCBJ. In the low bias voltage measurements ($V_b \approx 10$~mV)
the conductance of the contacts increases monotonically  while the
electrodes are continuously pushed together by the piezo driver. At
elevated bias voltages the conductance grows only to a certain value
(determined mainly by the magnitude of $V_b$) and thereafter
saturates at a constant level during subsequent electrode approaches
as large as 20~nm. This part of the conductance curve, where the
fluctuations  are caused by  temperature and current induced atomic
rearrangements, can be analyzed with histograms based on a $single$
sweep of the electrode approach. Furthermore, with fast data
acquisition one can get an insight into the dynamics of thermally
activated point contacts.

A typical conductance curve for a Pb MCBJ recorded at an approach
rate of approximately 0.01~nm/s and a bias voltage of 180~mV is
presented in Fig.~1. After the transition from tunneling to direct
contact the conductance of the connective neck increases reaching
$G\;\approx\;250 - 300\;G_0$ and subsequently remains nearly
constant (or rises very slowly) during further electrode approaches
over a distance of $\sim$~22~nm. Retracting the electrodes back to
the tunneling range reveals an increase of $\sim$~24~nm in electrode
separation with respect to their initial position. This indicates
irreversible changes in the electrode relief (reduction of electrode
length) caused by continuous transfer of atoms from the neck area to
more remote parts of the electrode. In our opinion two mechanisms
are responsible for the effect observed.

(i). Surface diffusion of atoms: The high level of power dissipation
(up to $10^{-3}$~W) in the region with characteristic dimension of
$L_i$ around the contact raises  the neck temperature and therefore
the atomic diffusion. This induces motion of atoms from the ``hot
spot'' in the center of the connective neck to colder areas of the
electrodes.

(ii). Electromigration of atoms: The estimated current density in
the center of the point-contact presented in Fig.~1 is
$j\,\sim\,10^{10}$~A/cm$^2$. At such current densities the electron
wind forces resulting from the scattering of electrons on lattice
defects are  $F_{ew}\,\approx\,0.1 - 0.2$~nN. \cite{yas,schm} These
current-induced forces are one order of magnitude too small to break
bonds between atoms but are able to enhance diffusion and make
atomic flux directional causing a preferential transfer of atoms to
one of the electrodes. Breaking of gold nanowires caused by
electromigration of atoms at $V_b\;\approx\; 300\;-\;500$~mV was
reported recently by Park $et\; al.$ \cite{park}

The neck conductance remains nearly constant because of a peculiar
``feedback'' of natural origin. In a voltage biased measurement the
increase of contact diameter in the course of electrode approach
leads to an increase  of $G$ and power dissipation. At $L_i \gtrsim
d$ the volume where this power is dissipated remains the same ($\sim
L_i^3$), thus the temperature increases, causing an exponential
increase of diffusion. Thus, the number of atoms leaving the neck
area per unit of time increases and the diameter is reduced back.

At higher bias ($V_b\approx 250-300$~mV) the contact conductance
saturates already at $40-80~G_0$. An example of the flat part for
such a $G(z)$ dependence is presented in Fig.~2a (with no more than
5\% of the experimental points plotted) and a fragment of this curve
is shown on an extended scale in Fig.~2b. The random changes in the
conductance are caused by a continuous rearrangement of atoms in the
neck area. Halting of the electrode approach eventually reduces the
amplitude of $G$ variations by a factor of $\approx 3-10$. The mean
value of $G$ remains practically the same or slightly decreases. The
conductance does not relax to a lower value since at extremely slow
approach rate the system is already in a quasi-equilibrium state. An
increase of bias voltage for stationary contact reduces the
conductance, whereas a decrease leaves the conductance unchanged as
no diffusion to the overheated neck area from the colder part of
electrodes can occur. (The former effect imposes a limitation on
measurements of current-voltage dependencies. \cite{IV}) It should
be noted, though, that power dissipation in this contact
(proportional to $V_b^2 \cdot G$) is $\approx$ 2 times less than in
the previous case (Fig.~1), whereas current density increases.
Although it is very difficult to make any reasonable estimation for
the  connective neck temperature in this intermediate ``diffusive''
range, it may be safely suggested that the contribution of the
current-induced processes in neck evolution increases with
increasing bias voltage.

The conductance histogram for this scan is presented in (Fig.~3a)
and is nearly gaussian shaped. It is slightly cut at the low
conductance end and extended to larger $G$.  The histogram plot for
the {\it difference} in conductance between the neighboring data
points $\Delta G \,=\, G_{n+1} - G_n$ for the same scan is shown in
Fig.~3b. The significance of this type of histogram for data
analysis will become clear later. At this point it simply
demonstrates that on the time scale of the data acquisition (1~ms)
no preferential values of $\Delta G$ are observed. As the
conductance of a point contact is mainly determined by a volume of
$d^3$ around the contact centre, hundreds of atoms are responsible
for the resistance of this junction. The classical continuous
spectrum demonstrated by both figures is not surprising, as the
number of different metastable atomic configurations is so large,
that they can hardly be resolved as distinct peaks in conductance
histograms. Or speaking in terms of conductance channels, for
contacts of the above size the spacing between the energies of
different channels is so small that it can easily be smeared either
by temperature or by the enhanced scattering inside the contact.

At $V_b$ as high as 300 -- 400 mV the neck conductance is restricted
to 10-20 quantum units  over 10 -- 15 nm of electrode approach. In
this case the power dissipated in the electrodes, is 5 to 10 times
less than that for point-contact presented in Fig.~1. Moreover,
contacts of such dimension ($\sim$~1 nm) are quasi-ballistic, thus
no significant power dissipation takes place in the nearest vicinity
of connective neck. At the same time the current density nears
$10^{11}$~A/cm$^2$, greatly increasing the contribution of
electromigration and current-enhanced diffusion to the neck
evolution. Typical $G(z)$ dependencies are presented in Fig.~4 for
Al and Pb MCBJ. Again only 1 -- 3\% of all recorded points are
plotted for the sake of better clarity, as the total number of
different neck configurations measured per one scan reaches
$\sim\,10^5$. Both curves show a pronounced step-like structure and
clear cut recurring jumps of neck conductance between two or more
levels. In the following we concentrate on these phenomena.

Just as usual conductance curves these $G(z)$ dependencies are
inherently irreproducible for different samples since their behavior
is determined by fine details of ill-controllable surface relief of
the electrodes. The general shape of $G(z)$ curves does not depend
appreciably on the approach rate $S$ in the 5 to 50 pm/s range. An
increase of $S$ above 100 pm/s produces effects similar to a
decrease of $V_b$. At $S\,\gtrsim \,1$ nm/s conductance curves
assume the standard low bias behavior: monotonous increase in $G$ up
to $10^4\;G_0$. This shows that in reality the process of atomic
rearrangements in the connective neck (even at elevated temperature)
could be many orders of magnitude slower than it was suggested in
molecular dynamic simulations. \cite{brat} Thus, it is hardly
surprising that  in experiments with gold relays \cite{yas} at
approach rates of $\gtrsim \; 10^5$~nm/s the authors did not find any
difference between conductance histograms taken at low $V_b$ and
bias voltage as high as 500~mV and observed a plateau in conductance
curves corresponding to the one-atom contacts even at $V_b\;=$~1.5
V. Contrary to this, in experiments with Au (as well as Al and Pb)
MCBJ we were unable to create one-atom contact already at
$V_b\;\geq\;0.8 - 1.0 V$ and $S\;\sim$ 0.01 nm due to field
desorption or field evaporation of the surface atoms at small
interelectrode distances.

Figure 5 presents a histogram plot for Al MCBJ based on more than
$10^5$ of different connective neck configurations recorded in a
single 20 nm scan. It displays a number of well-defined peaks in the
range from 0.5 to 18 $G_0$. The total conductance of a contact with
few atoms is determined by the sum of $N$ independent conducting
channels with their respective transmission coefficients
$G\,=\,G_0\sum_{i=1}^N \tau_i$. Rearrangement of atomic positions
may result in a completely different transmission set $\top =
\sum_{i=1}^N \tau_i$ for the whole neck. \cite{scheer} Data
presented in Fig.~5 clearly show that in the course of continuous
neck evolution under elevated temperature and high current density
the selection of the most stable neck geometries occurs.  Only the
preferable (energetically favorable) atomic arrangements can survive
for a long enough time to give a rise to the sharp peaks in the
histogram.

In Al and Pb contacts the $G(z)$ dependencies are especially
sensitive to the elastic deformation of the connective neck in the
process of creation and break of contacts at low bias voltage.
\cite{alpb,kob} This is reflected as a steep slope of  plateaus in
the conductance staircase and causes a considerable smearing of the
peaks in the histograms. An example of a conventional histogram for
Al based on 5000 individual $G(z)$ traces measured at low bias
voltage is presented in Fig.~6. (We did not find any perceptible
difference in peak positions for histograms taken at low bias
voltages and at 300 mV. This indicates a rather small non-linearity
of I-V characteristics of a few atom Al MCBJ). At high $V_b$ and a
slow approach of electrodes the deformation of the neck is much
moderate because of a fast relaxation of strain through diffusion of
atoms and electromigration of defects. For this reason the plateaus
in our conductance curves (Fig.4) are tilted less than in $G(z)$
curves recorded at low bias \cite{alpb}.

We performed further analysis of the dynamic evolution of contacts
with conductance restricted by 20 -- 25 quantum units using selected
fragments typical for most of the conductance curves. An example is
presented in Fig.~7a. The contact conductance switches between a few
preferential levels, though we did not find any regular trend in the
peak positions in the traditional histogram plot (Fig.~7c). At the
same time the histogram for the difference of conductance between
neighboring data points $\Delta G \, = \,|G_{n+1}(z) - G_n(z)|$
demonstrates the succession of 4 - 5 maxima separated by 0.7 $G_0$
(Fig.~7d). Since the value of 0.7 $G_0$ is very close to the
experimentally measured \cite{thesis} and calculated \cite{alpb}
conductivity through a one-atom Al contact for an undistorted
lattice, it is reasonable to relate this unexpected periodicity to
variations in the number of atoms in the connective cross section by
$\pm 1, .. \pm 5$ during the time between successive measurements
(1ms). However, in traditional histograms, based on thousands of
conductance traces no such periodicity was observed, and the
separation between  peaks is much larger (1.1 $G_0$). This means that
the conductance of a single neck with n atoms in the cross section
is not simply the sum of n one-atom contacts. The observed effect is
most likely related to momentary formation and breaking of  one-atom
contacts isolated from the main neck.  Due to surface roughness at
the atomic scale two electrodes can be bridged for a short time by
mobile atoms moving along the surface (see model in Fig.~7b). The
conductance of parallel connected one-atom contacts separated both
from each other and from the main neck is simply given as a sum of
the conductance of the individual contacts, which explains the
periodicity of the peaks in the histogram plot in Fig.~7d. It should
be emphasized that the visual appearance of this histogram plot
probably depends on the recording rate: an increase in this will
leave us finally with a sole peak at 0.7 $G_0$ corresponding to the
creation or break of a single one-atom contact between two
successive measurements. At the same time the possibility of
simultaneous formation of two or more one-atom contacts in a single
event cannot be ruled out. This result opens an interesting
possibility to study the dynamics of single atom contact formation
under different circumstances by extending the time scale of data
recording into microseconds range.

Yet another two typical fragments of conductance curves for Pb and
Al MCBJ with somewhat lower conductance are presented in Fig.~8.
They exhibit typical random telegraph noise behavior observed on
many occasions in different mesoscopic systems. In our case these
fluctuations between discrete levels are caused by switching of Al
or Pb atoms between metastable positions. In both cases the
repeatable jumps with larger amplitude ($0.8~G_0$ for Al, and
$2.4~G_0$ for Pb) might correspond to the formation and break of
isolated one-atom contacts (the calculated conductivity through a
one-atom Pb contact is $\approx$ 2.5 $G_0$ \cite{scheernat,alpb}).
The above transitions are alternating with conductance jumps of
lesser amplitude: $\Delta G=1.1~G_0$ for Pb and $\Delta G=0.5~G_0$
and $0.2~G_0$ for Al. These additional two-level transitions can be
explained by the recurring motion of atoms comprising the one-atom
contact, which only slightly alternates their relative positions.
The conducting channels in single atom contacts of polyvalent metals
are generally associated with atomic orbitals.\cite{alpb} In the
case of different orientation and spatial extent of orbitals (which
is not improbable for atoms at a metal-vacuum interface) a small
displacement of atoms can change the overlapping of electron
densities associated with different orbitals and therefore induce
the complete shut down of certain channels. For instance the
transition with $\Delta G $ of 0.5 and 0.2 $G_0$ in Al may be linked
to the switching off and on of high transmissive $sp_z$ or low
transmissive $p_{x,y}$ channels, respectively, while the transitions
with $\Delta G \approx 1.1\,G_0$ in Pb are  close to conductance
through one of the three conductive channels. \cite{scheernat,alpb}.
Thus, the transitions of smaller amplitude in Fig.~8 might be
related to changes in the $number$ of conducting channels through
parallel-connected one-atom contact.

To be certain that the above effects are not related only to
selected fragments of conductance curves we build up differential
histograms for complete scans with a large number ($\sim 10^5$) of
different connective neck configurations. These histograms (Fig.~9)
for Al MCBJ show shoulders or smeared peaks at $0.6-0.8~G_0$, around
$1.5~G_0$ and sometimes even at $2.0-2.2~G_0$ corresponding to the
formation and break of one or more single-atom contacts separated
from the main neck. A comparison to Fig.~7 reveals an additional
peak at $\approx$ 1.1 $G_0$ which means that transitions with
$\Delta G$ = 1.1 $G_0$ are also preferable. Analysis of the
traditional conduction histograms (as the one presented in Fig.~6)
shows that the separation between two neighboring  peaks (or
statistically more favorable neck configurations) is indeed very
close to 1.1 $G_0$ (this fact was mentioned in Ref.\
\onlinecite{thesis} as well). This is further supported by the peak
at $1.1~G_0$ in the differential histogram for the same data set of
5000 individual conductance curves (see inset in Fig.~6). Although
the reason for nearly equidistant spacing of peaks in Al conductance
histograms is not yet clear, one can state that the conductance
difference between the stable configurations of a small neck is
approximately $1.1~G_0$. The peak at $1.1~G_0$ in Fig.~9 might also
reflect the switching between the stable geometries of the main neck,
while the rest of the peaks come from the formation and breaking of
isolated one-atom contacts and not expected in the traditional low
bias voltage histogram.

To summarize briefly, in the connective necks  with conductivity in
the range below 20 $G_0$ the superposition of the following effects
have been seen:

- switching between the favorable neck configurations (leading to a
$\Delta G$ = 1.1 $G_0$ peak in differential histograms for Al MCBJ);

- connecting and disconnecting of one-atom contacts separated from
the main neck (which are resulting in a conductivity jump of 0.7
$G_0$ in Al and 2.5 $G_0$ in Pb MCBJ);

- switching "on" and "off" the individual conductive channels
associated with atomic orbitals in the one-atom contacts
(conductance jumps on a smaller scale).

The increase of $V_b$ above 400 $mV$ results in unstable behavior of
Al contacts with frequent disconnection  of electrodes (jumps to
tunneling and back). On many occasions we were able to maintain
contacts of one atom over the approach distance up to 2-3 nm. The
histogram for such conductance curves mostly reveals ``fine
structure'' of the first peak - splitting into two clearly cut
components (Fig.~10a). To explain this effect one may consider
switching of the one-atom neck between two different well defined
and stable configuration which are incorporating the adjacent atoms
of both electrodes  (e.g. bent and straight necks connecting two Al
electrodes are supposed to have a different conductance \cite{kob}).
In conventional low bias conductance histograms for Al the first
peak is rather broad and centered around 0.8 $G_0$. However we found
that on rare occasions it has a low-energy "shoulder" or can even be
split (see Fig.~10b), indicating that the contact break may occur in
two different ways.

Concluding, we used a new method for investigation of connective
neck evolution. It is based on the fact that at elevated bias
voltages and slow approach of the electrodes the conductance of the
constriction remains within a limited range over a relatively long
distance of electrode displacement due to the current-enhanced
diffusion and electromigration of atoms out of the overheated area.
The dynamics of neck evolution can be traced using $G(z)$
dependencies and a differential histogram technique. For
conductances higher than $\sim$ 50 $G_0$ the fluctuation of $G(z)$
demonstrates a quasi-classical continuous spectrum. In the range up
to 20 fundamental units the quantum nature of conductance shows up
in a discrete spectrum due to an ever changing number of individual
conductance channels, continuous formation and rupture of parallel
connected one-atom contacts isolated from the main neck and
transition between preferable neck geometries. The dependence of the
$G(z)$ traces on the approach rate shows that the relaxation of
atoms in the connective neck occurs much slower than it had been
suspected before.

\section*{Acknowledgments}

We are indebted to I.K. Yanson for stimulating discussions. Part of
this work was supported by the Stichting voor Fundamenteel Onderzoek
der Materie (FOM) which is financially supported by the Netherlands
Organization for Scientific research (NWO), the Hungarian Research
Funds OTKA T026327 and N31769 and a NWO grant for Dutch-Hungarian
cooperation. O.I.S. wishes to acknowledge the NWO for a visitor's
grant.

\begin{figure}
\centering
\includegraphics{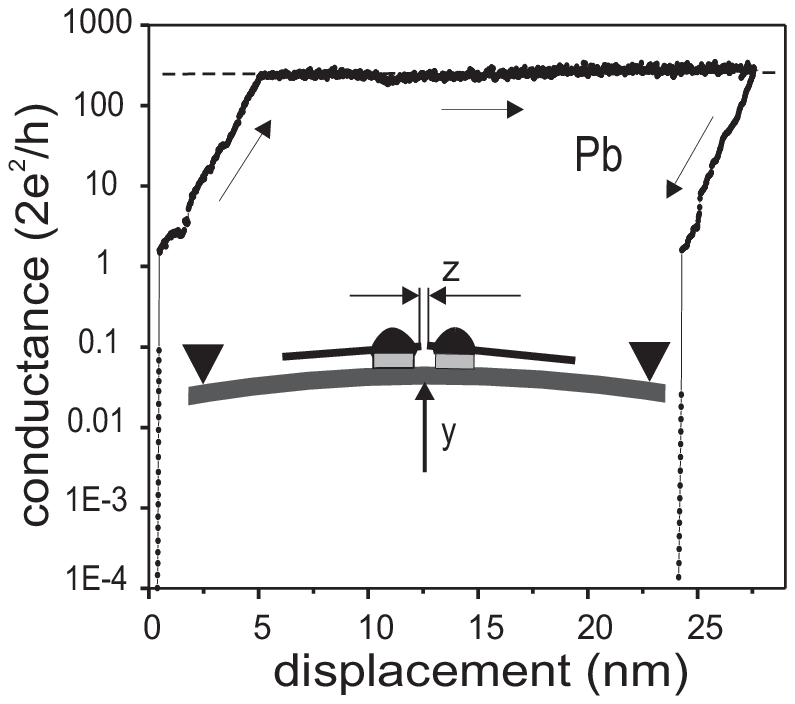}
\vspace{10truemm}
\caption{The conductance of a Pb MCBJ as a function of
electrode displacement at a bias voltage of 180 mV. The inset shows
the design and the principle of a mechanically controllable break
junction  (see also text).  A bending force causes a vertical
displacement $y$, which leads to a displacement $z$ between the
electrodes. For clarity, all distances have been exaggerated.}
\end{figure}

\newpage

\begin{figure}
\centering
\includegraphics{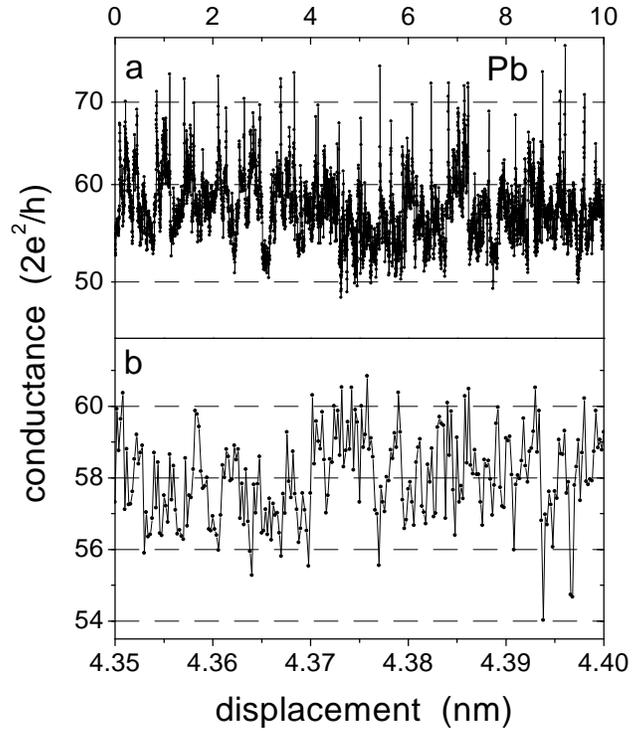}
\vspace{10truemm}
\caption{a) $G(z)$ dependence for Pb MCBJ recorded at $V_b$ = 280 mV
and approach rate $S$ = 40 pm/s (only 5\% of measured points are
plotted). b) A part of the $G(z)$ trace in the upper panel presented
on an extended scale.}
\end{figure}

\newpage

\begin{figure}
\centering
\includegraphics{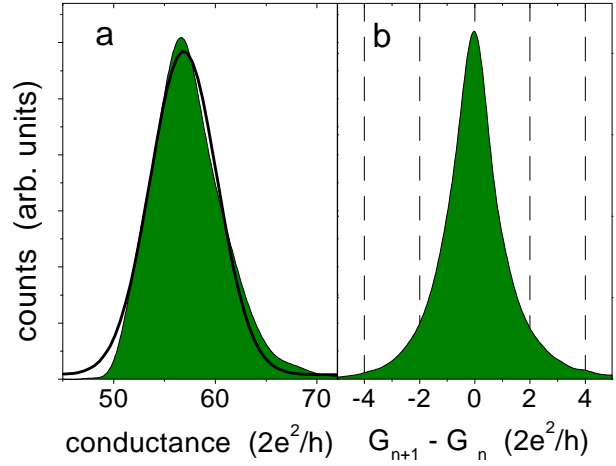}
\vspace{10truemm}
\caption{a) Conductance histogram and gaussian fit (solid line) for
the $G(z)$ curve presented in Fig.~~2a. b) Histogram plot of
conductance difference between neighboring points $G_{n+1}-G_{n}$
for the same curve.}
\end{figure}

\newpage

\begin{figure}
\centering
\includegraphics{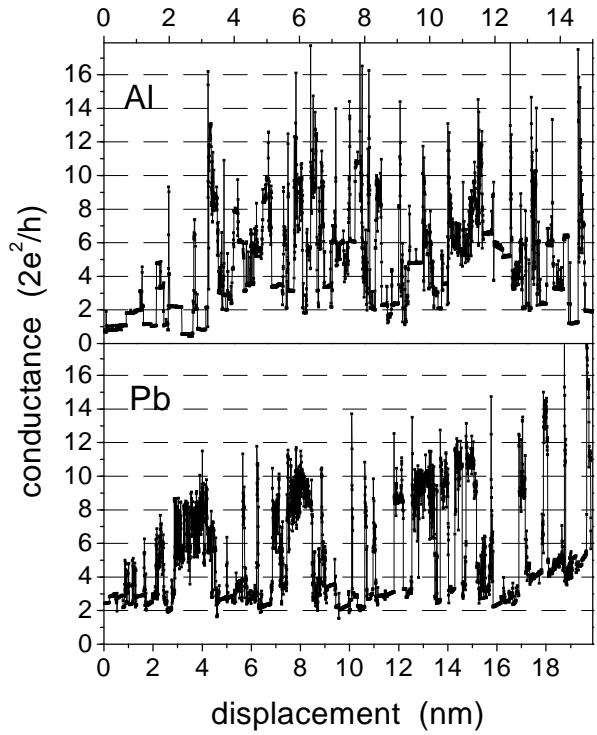}
\vspace{10truemm}
\caption{$G(z)$ dependencies for Al  (recorded at $V_b$ = 340 mV and
$S$ = 40 pm/s) and  for Pb (recorded at $V_b$ = 360 mV and $S$ = 30
pm/s) in quantum regime of conductance.}
\end{figure}

\newpage

\begin{figure}
\centering
\includegraphics{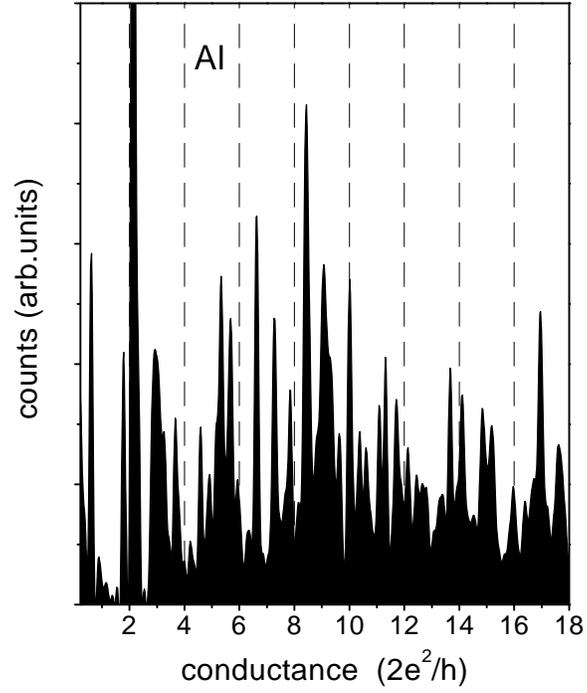}
\vspace{10truemm}
\caption{Conductance histogram for Al corresponding to more than
$10^5$ individual neck configurations. The raw data were smoothed
with a narrow window of 0.1 $G_0$. The original $G(z)$ curve was
recorded at $V_b$ = 350 mV and $S$ = 30 pm/s.}
\end{figure}

\newpage

\begin{figure}
\centering
\includegraphics{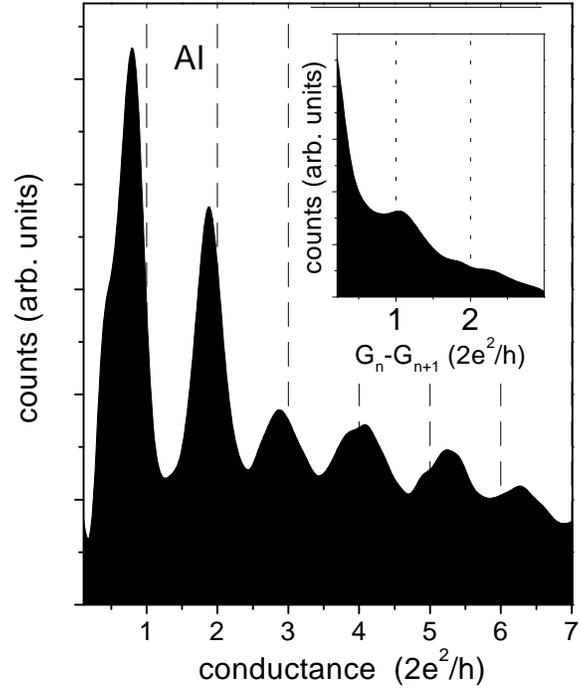}
\vspace{10truemm}
\caption{Low bias conductance histogram for Al MCBJ based on 5000
individual traces recorded at 100 mV bias. Inset: differential
histogram for the same set of data displaying a peak around 1.1
$G_0$.}
\end{figure}

\newpage

\begin{figure}
\centering
\includegraphics[width=7truecm]{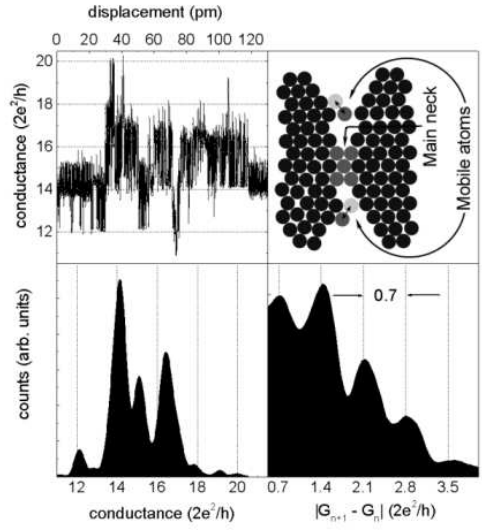}
\vspace{10truemm}
\caption{(a) A fragment of a $G(z)$ dependence for Al ($V_b$ = 340 mV
and $S$ = 50 pm/s); (b) model of formation and break of the single
atom contacts being separated from the main neck and from each
other; (c) Histogram  plot for the conductance curve (a); (d)
histogram plot for the absolute value of the conductance difference
$\Delta G$ between neighboring points of curve (a) demonstrating a
succession of maxima separated by $\sim 0.7\;G_0$}
\end{figure}

\newpage

\begin{figure}
\centering
\includegraphics{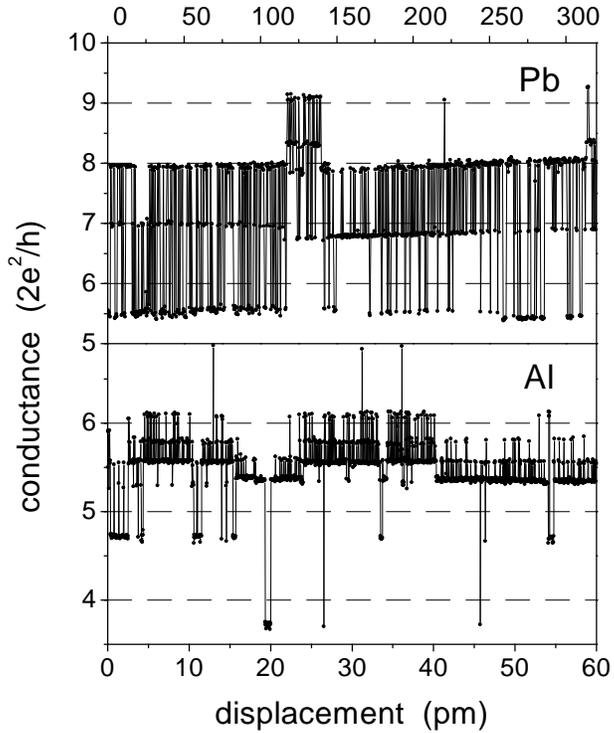}
\vspace{10truemm}
\caption{Short fragments of conductance curves for Al ($V_b$ =
350 mV and $S$ = 30 pm/s) and Pb $(V_b$ = 360 mV and $s$ = 30 pm/s)
demonstrating random telegraph noise patterns corresponding to
fluctuation of atoms between two metastable states accompanied by
creation (rupture) of one-atom point-contacts. Conductance jumps of
smaller amplitude corresponds to changes in the number of conducting
channels due to incomplete overlap of orbitals.}
\end{figure}

\newpage

\begin{figure}
\centering
\includegraphics{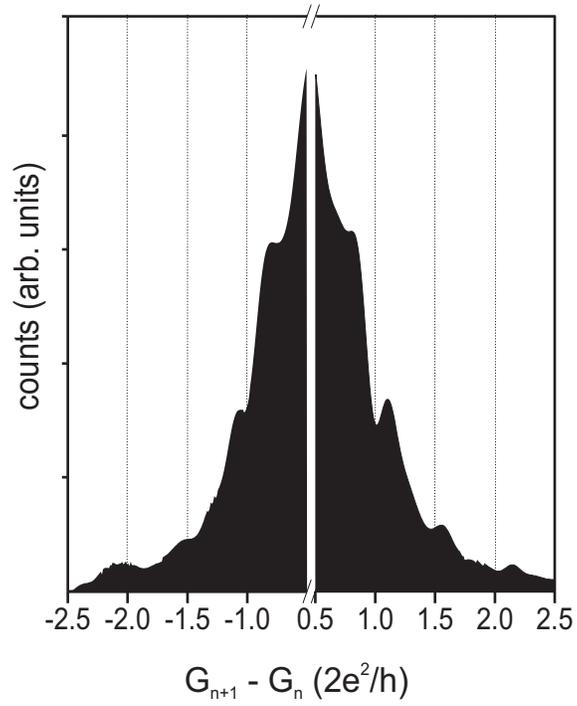}
\vspace{10truemm}
\caption{Differential histogram for Al MCBJ demonstrating
singularities caused by formation and break of one-atom contacts and
by transitions  between the stable configurations of the neck (see
text).}
\end{figure}

\newpage

\begin{figure}
\centering
\includegraphics{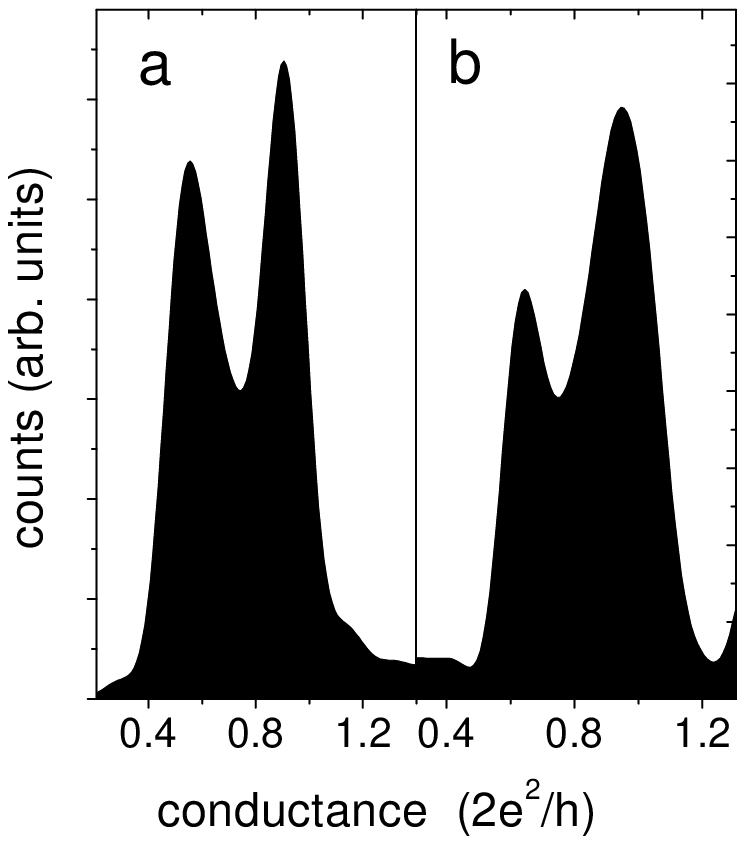}
\vspace{10truemm}
\caption{``Fine structure'' of the first peak in conductance
histograms for Al  one-atom contacts. a) Histogram made during
continuous approach ($V_b$ = 400 mV and $S$ = 30 pm/s) b)
Conventional histogram based on 5000 $G(z)$ curves recorded at 100
mV bias.}
\end{figure}


\begin{references}
\bibitem{IV} Typically the upper voltage limit of $I-V$ curves for Al MCBJ,
attainable  without irreversible changes in conductivity, is
$\lesssim$ 200 mV for contact with $G \approx$ 100 - 200 $G_0$. It
increases up to 250 - 300 mV for $G \approx$ 30 - 100 $G_0$, reaches
350 - 400 mV for G in the range 10 - 30 $G_0$ and peaks at 500 - 600
mV for the stable contact configurations with conductivity less than
10 $G_0$. In all cases the relative change of conductivity at the
highest $V_b$ with respect to zero bias conductivity was $\lesssim$
3 - 5 \% and therefore the non-linearity of $I-V$ curves has little
effect on our data.
\end{references}
\end{document}